\documentclass[preprint2]{aastex}
\usepackage{amsmath}
\usepackage{amssymb}
\usepackage{graphicx}

\def\aap{A\&A} 
\def\apj{ApJ} 
\def\apjl{ApJL} 
\def\apjs{ApJS} 
\def\mnras{MNRAS} 
\def\nat{Nature} 

\def\eg{{e.g.}} 
\def\ie{{i.e.}} 
\def\etc{{etc}} 

\def\GHz{{\rm GHz}} 
\def\m{{\rm m}} 
\def\mum{\mu\m} 
\def\cm{{\rm c}\m} 
\def\muas{\mu{\rm as}} 

\def\SgrA{{Sgr A*}}

\title{Frequency-Dependent Shift in the Image Centroid of the Black
  Hole at the Galactic Center as a Test of General Relativity}

\author{Avery E. Broderick\altaffilmark{1} and Abraham Loeb\altaffilmark{2}}
\affil{Institute for Theory and Computation, Harvard University, CfA,
  MS 51, 60 Garden Street, Cambridge, MA 02138, USA}
\altaffiltext{1}{abroderick@cfa.harvard.edu}
\altaffiltext{2}{aloeb@cfa.harvard.edu}

\shorttitle{\SgrA~Centroid Shift}
\shortauthors{A.~E. Broderick \& A. Loeb}

\begin{document}

\begin{abstract}
The inferred black hole in the Galactic center spans the largest
angle on the sky among all known black holes.  Forthcoming observational
programs plan to localize or potentially resolve the image of \SgrA~to an
exquisite precision, comparable to the scale of the black hole horizon.
Here we show that the location of the image centroid of \SgrA~should depend
on observing frequency because of relativistic and radiative transfer
effects.  The same effects introduce a generic dependence of the source
polarization on frequency.  Future detection of the predicted centroid
shift and the polarization dependence on frequency can be used to determine
the unknown black hole spin and verify the validity of General Relativity.
\end{abstract}


\maketitle

\section{Introduction}

The individual orbits of fast-moving stars near the dynamical center of the
Milky Way galaxy indicate the existence of a black hole of mass $\sim
4\times 10^6$ solar masses at the position of the radio/infrared source
\SgrA~\citep{Scho_et_al:03,Ghez_et_al:05}.  The angular scale occupied by
the horizon of this black hole, $5$--$10$ micro-arcseconds ($\mu$as), is a
factor of $\sim2$ larger than the nuclear black hole in the distant galaxy
M87 and orders of magnitude larger than all other known black holes.
Recent observations revealed infrared emission from the radio source
\SgrA~which flares on a timescale of tens of minutes, comparable to the
horizon crossing time of accreting gas \citep{Genz_etal:03}. This data has
motivated an observational program to monitor shifts of tens of $\mu$as in
the centroid of the infrared image of \SgrA~due to an orbiting hot spot in
the accretion flow during flares \citep{Brod-Loeb:05}, using the newly
constructed PRIMA instrument on Very Large Telescope (VLT)
\citep{Paum_etal:05}. In parallel, there are plans to image \SgrA~at $\sim
20\mu$as resolution with a Very Large Baseline Array (VLBA) of
observatories at sub-millimeter wavelengths \citep{Miyo_etal:04}, at which
the accreting gas is expected to be optically-thin and the black hole
shadow may be recovered \citep{Falc-Meli-Agol:00}.

The faint emission from \SgrA~is commonly interpreted as a consequence of a
radiatively inefficient accretion flow around the black hole
\citep{Yuan-Quat-Nara:03}.  The observed break in the sub-millimeter
spectrum is thought to signal the transition of the flow to the optically
thin regime at wavelengths below the break.  The flux below the break
wavelength is expected to be dominated by emission from the vicinity of the
black hole horizon, where general relativistic effects become significant.
These effects result from the warping of spacetime by the black hole (\eg,
gravitational redshift, rotation of polarization via parallel propagation)
and the rapid motion of the disk near the {\it innermost stable circular
orbit} (ISCO; \eg, Doppler shift, beaming, Lorentz boosted polarization
vectors, \etc.).  Although attempts have been made to incorporate these
effects into the determination of the unpolarized spectrum in the optically
thin limit \citep{Kurp-Jaro:00}, earlier studies have not included a
complete treatment of the polarization spectrum\footnote{At frequencies
significantly larger than the plasma and cyclotron frequencies, treatments
that assume the plasma eigenmodes are uncoupled \citep[\eg][]{Brom-Meli-Liu:01}.
are inappropriate.  In the vacuum limit, the two plasma eigenmodes remain
strongly coupled such that the polarization is parallely propagated
regardless of the details of the plasma.} and are not applicable to the
sub-millimeter wavelength regime where the opacity is non-negligible.

Here we produce images for three variations on the model of
\citet{Yuan-Quat-Nara:03}, listed in Table \ref{models}, each adjusted to
roughly fit the observed radio and infrared spectrum.  The primary
distinction between the different cases in Table \ref{models} is the black
hole spin, ranging from non-rotating to maximally rotating ($a=0.998$).
This was done using the methods employed by \citet{Brod-Loeb:05} and
detailed in \citet{Brod-Blan:03,Brod-Blan:04}, and we direct the reader
to these papers for more information.  From the calculated images we
compute the unpolarized and polarized spectra as well as the location of
the image centroid as functions of frequency.

\section{Accretion Models}
\begin{deluxetable}{ccccc}
\tablewidth{0pt}
\tablecaption{\label{models}
Accretion Model Parameters}
\tablehead{
\colhead{$a\,(M)$} &
\colhead{$n_e^0\,(\cm^{-3})$} &
\colhead{$T_e^0\,(K)$} &
\colhead{$n_{\rm nth}^0\,(\cm^{-3})$} &
\colhead{$p_{\rm nth}$}}
\startdata
$0$ & $3\times10^7$ & $1.7\times10^{11}$ &
$8\times10^4$ & $-2.9$\\
$0.5$ & $3\times10^7$ & $1.4\times10^{11}$ &
$5\times10^4$ & $-2.8$\\
$0.998$ & $1\times10^7$ & $1.5\times10^{11}$ &
$1\times10^5$ & $-2.8$\\
\enddata
\tablenotetext{a}{In the following figures, the short-dash, long-dash
  and solid lines correspond to the $a=0$, $a=0.5$ and $a=0.998$
  models, respectively.}
\end{deluxetable}
We model the structure of the accretion flow based on the results of
\citet{Yuan-Quat-Nara:03}, who showed that the vertically-averaged
electron density and temperature could have a nearly power-law dependence
on radius.  Here we write the density, $n_e$, and temperature, $T_e$, of
the thermal electrons, as well as the density, $n_{\rm nth}$, of
non-thermal electrons, as having the spatial distribution
\begin{align}
n_e &= n^0_e \left(\frac{\rho}{M}\right)^{-1.1} \exp(-z^2/2\rho^2)
\nonumber\\
T_e &= T^0_e \left(\frac{r}{M}\right)^{-0.84}
\\
n_{\rm nth} &= n^0_{\rm nth} \left(\frac{\rho}{M}\right)^{p_{\rm nth}}
\exp(-z^2/2\rho^2) \,,
\nonumber
\end{align}
where $\rho$ is the cylindrical radius relative to the black hole spin
axis, and the assumed constants are listed in Table \ref{models} for three
fiducial black hole spins.  In all models, the non-thermal electrons have a
spectral index of 1.25 and a minimum Lorentz factor of 100. The magnetic
field strength is set to be a fixed fraction ($30\%$) of equipartition
relative to the protons and is chosen to be toroidal, as found in recent
general-relativistic magnetohydrodynamic simulations
\citep{DeVi-Hawl-Krol:03}.  The accreting gas is assumed to be in free fall
inside of the ISCO ($6M$, $4.233M$ and $1.237M$ for $a=0$, $0.5$ and
$0.998$, respectively), and in Keplerian rotation otherwise.  In all cases
the disk angular momentum was aligned with the spin of the black hole.

\section{Discussion}
\begin{figure*}
\begin{center}
\includegraphics[width=\textwidth]{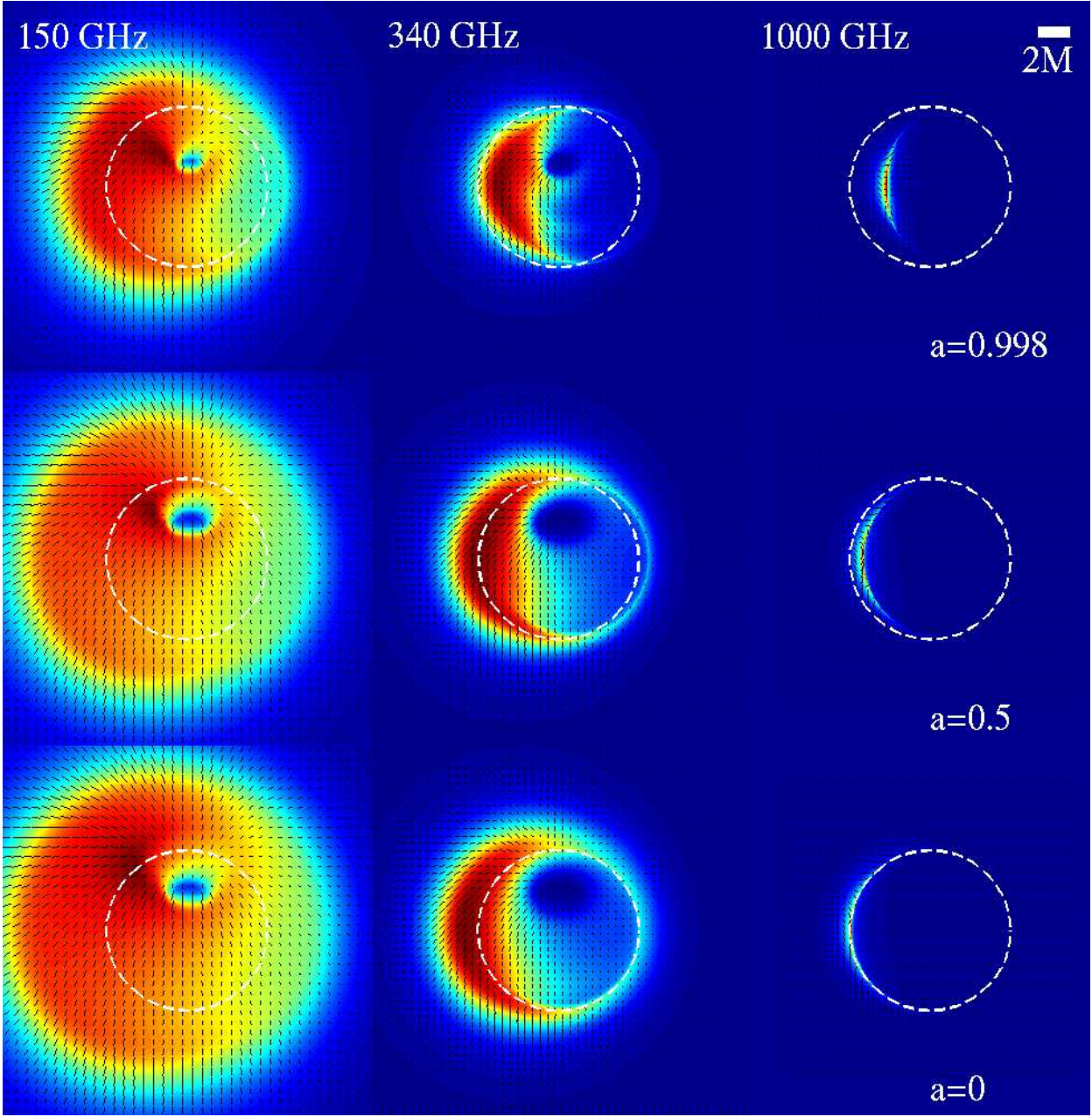}
\end{center}
\caption{Images of an accretion disk around a non-rotating (bottom),
moderately-rotating (middle), and maximally-rotating (top) black hole at
$150\,\GHz$ (left), $340\,\GHz$ (middle) and $1000\,\GHz$ (right) viewed at
an angle of $45^\circ$ relative to the spin axis.  Polarization tickmarks
are overlayed, the lengths of which are indicative of the amount of
polarized intensity ({\em not} polarization fraction).  The bar in the
upper-right corner indicates the scale $2M$ ($\sim10\muas$).  For
reference, the photon capture radius of a non-rotating black hole is
delineated by the dashed white line.  The intensity scale (red being bright
and blue being faint) is normalized separately for each image and the
vertical axis is aligned with the black hole spin.  The increasing
asymmetry in the images is primarily due to the special relativistic
effects, which become more significant at the ISCO for a rapidly rotating
black hole.  The dim region near the center of the $150\,\GHz$ images is
due to the funnel region of the thick disk, {\em not} due to the black hole
``shadow''.  The high frequency emission, arising from closest to the black
hole, is significantly offset between $a=0$ and $a=0.998$.}
\label{fig1}
\end{figure*}
As shown in Figure \ref{fig1}, the photosphere becomes smaller at higher
frequencies as expected.  However, special relativistic aberration and the
Doppler effect make the photosphere asymmetric about the rotation axis of
the accretion flow.  At high frequencies, the emission is dominated by a
small region located on the approaching side of the disk.  This has two
immediate consequences: a high polarization at high frequencies (10-20\%),
and a definite shift in the image centroid as a function of black hole spin
due to strong gravitational lensing.  At frequencies $\ga 340\,\GHz$ the
black hole ``shadow'' discussed by \citet{Falc-Meli-Agol:00} \citep[see
also][]{Taka:04,Miyo_etal:04} becomes apparent.  However, in contrast to
previous work, the rotation of the disk makes the shadow highly asymmetric
and even for the non-rotating case it only roughly follows the photon
capture cross section (shown by the white dashed line).  Because rotation
is a generic feature of the accretion flow, it is unlikely that
sub-millimeter imaging of \SgrA~will show a well defined shadow, as
suggested before based on a much more simplistic model for emissivity
profile.

\begin{figure}
\begin{center}
\includegraphics[width=\columnwidth]{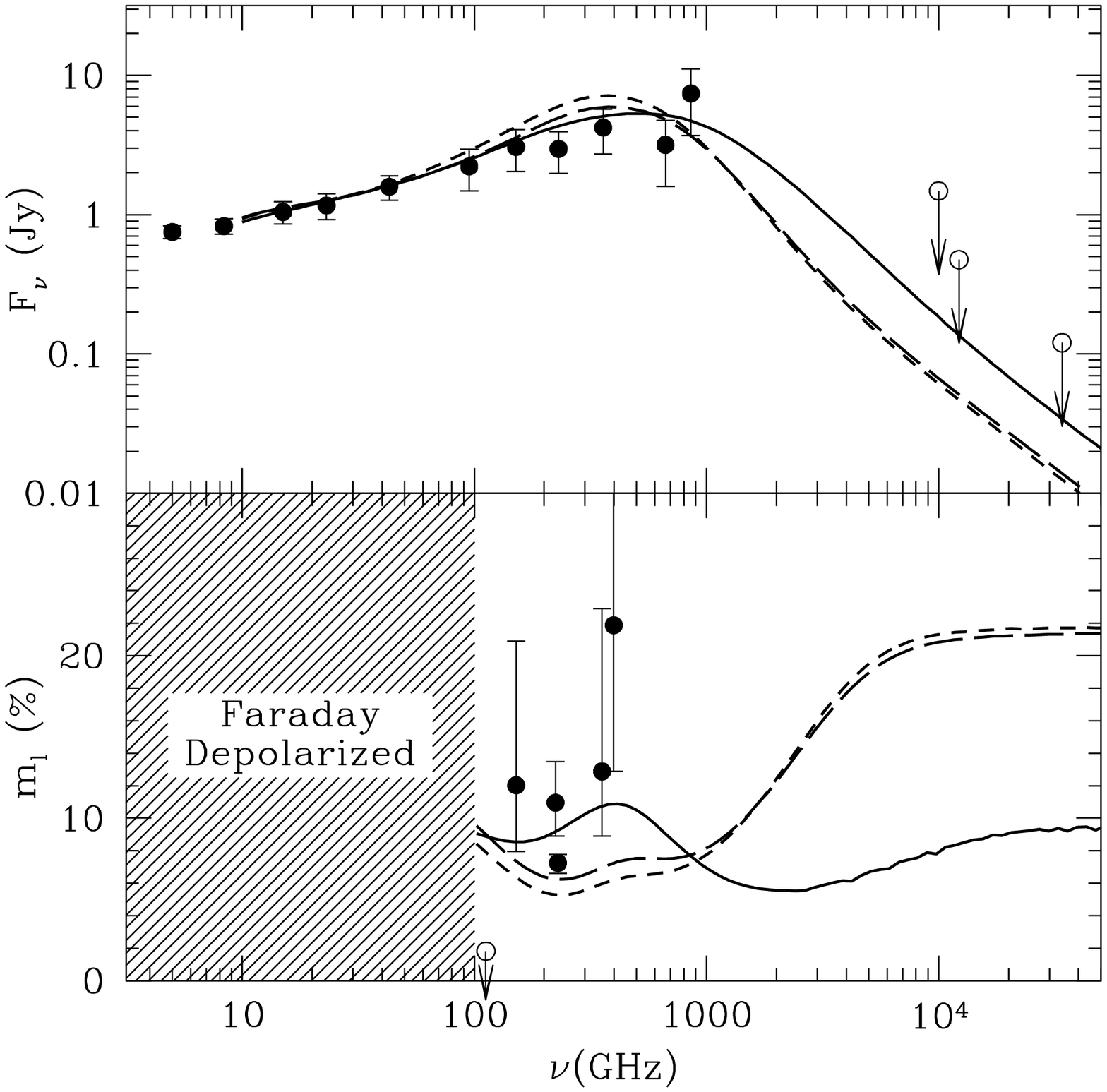}
\end{center}
\caption{The spectral flux (upper panel) and linear polarization fraction
(lower panel) are shown for the non-rotating ($a=0$, short-dash),
moderately rotating ($a=0.5$, long-dash) and maximally rotating ($a=0.998$,
solid) black hole.  The accretion disk models are listed in Table
\ref{models} and the disk is viewed at an angle of $45^\circ$.  In all
cases the spectra give an acceptable fit to the existing data with the
maximally rotating black hole model giving the best fit.  The polarization
fraction appears to be a sensitive diagnostic of the spin of the black hole
near $350\,\GHz$ and above $10^{4}\,\GHz$ ($30\,\mum$).  Below $100\,\GHz$
it is believed that \SgrA~is Faraday depolarized, and thus our calculation
(in which Faraday rotation was ignored) is inapplicable.  The data are
taken from
~\cite{Aitk_etal:00,Bowe-Falc-Saul-Back:02,Bowe-Wrig-Falc-Back:03}.  The
polarization at high frequencies is a direct result of the localization of
the emission evident in Figure \ref{fig1}.}
\label{fig2}
\end{figure}
The unpolarized and polarized spectra are shown in Figure \ref{fig2}.  High
black hole spin (and thus high Keplerian velocity at the disk inner edge)
leads to a broadened sub-millimeter bump, yielding a marginally better fit
to the data.  Given the current uncertainties inherent in modeling the
inner edge of the accretion flow, this is not necessarily evidence for
spin.  However, it does demonstrate the importance of relativistic effects
between $30$ and $300\,\GHz$.  The polarized spectra shows the anticipated
asymptotic behavior at high observing frequencies, which strongly
discriminates between low and high spin black holes.  In addition to the
high frequency behavior there is a sub-millimeter bump in the polarization
fraction as well.  Again the data appears to marginally favor high spin.
However, in this case as well, there is significant uncertainty in the
emission characteristics, most notably the geometry of the magnetic field.
Nevertheless, our assumption of a toroidal field is likely to be justified
in the inner regions of the disk due to the strong shear present in that
region.

\begin{figure}
\begin{center}
\includegraphics[width=\columnwidth]{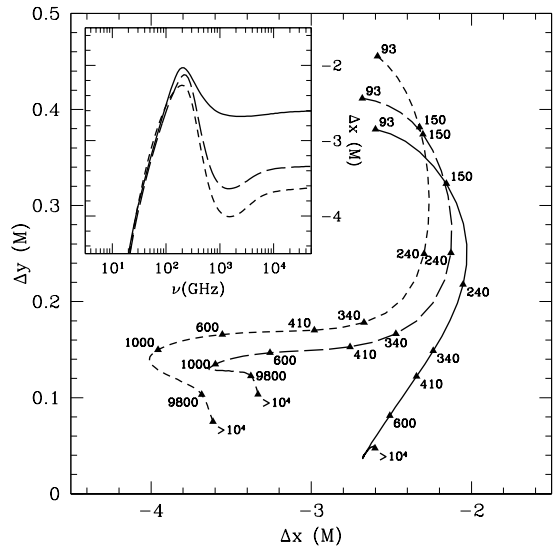}
\end{center}
\caption{The position of the centroid of the disk image is shown as a
function of observing frequency (labeled in $\GHz$) for the three models
listed in Table \ref{models} viewed at an angle of $45^\circ$.  The
position shifts are measured relative to the black hole (where the $y$-axis
is aligned with the black hole spin).  The projection of the trajectories
on the $\Delta x$ axis are shown as functions of frequency in the inset.
For \SgrA, the angular scale of the
axes is $M\simeq5\muas$ times the black hole mass in units of $4\times
10^6M_\odot$.}
\label{fig3}
\end{figure}
The position of the image centroids as a function of frequency is plotted
in Figure \ref{fig3}.  As suggested by Figure \ref{fig1} the image
centroids asymptote to differing fixed positions at high frequencies.  The
asymptotic positions are determined by strong gravitational lensing, and
thus are sensitive to the parameters of the black hole.  In contrast, when
the photosphere has a large radius (\ie, at low frequencies), the black
hole spin has little influence and thus the image centroids become
degenerate.  The centroid location rapidly changes as the photosphere
shrinks to the scale of the horizon.  This necessarily occurs during the
transition between the optically thick and the optically thin regimes.
Therefore, high precision absolute astrometry of the image centroids at
radio and infrared wavelengths can constrain the black hole spin.

While quantitative differences exist for different magnetic field
geometries in both the polarized spectra and the locations of the image
centroids, the qualitative characteristics (which depend only upon the
asymmetric opacity and relativistic effects) remain the same.  As a result,
detailed multi-wavelength studies of the polarization and image centroids
provide a method by which the spin of the black hole at the Galactic center
may be measured.

\acknowledgments
 This work was supported in part by grants from NSF and NASA.
  A.E.B. gratefully acknowledges the support of an ITC Fellowship from
  Harvard College Observatory.

\end{document}